\documentstyle[12pt]{article}
\newcommand{\extraspace}{\addtolength{\abovedisplayskip}{2mm}
                        \addtolength{\belowdisplayskip}{2mm}
                        \addtolength{\abovedisplayshortskip}{2mm}
                        \addtolength{\belowdisplayshortskip}{2mm}}
\newcommand{\be}{\begin{equation}\extraspace}

\newcommand{\ee}{\end{equation}}

\newcommand{\bea}{\begin{eqnarray}\extraspace}
\newcommand{\beastar}{\begin{eqnarray*}\extraspace}
\newcommand{\eea}{\end{eqnarray}}
\newcommand{\eeastar}{\end{eqnarray*}}

\newcommand{\nonu}{\nonumber \\[2mm]}

\newcommand{\half}{{\textstyle \frac{1}{2}}}
\newcommand{\quart}{{\textstyle \frac{1}{4}}}

\newcommand{\threehalves}{{\textstyle \frac{3}{2}}}

\newcommand{\del}{\partial}
\newcommand{\eg}{{\it e.g.}}
\newcommand{\ie}{{\it i.e.}}

\newcommand{\vac}{|0\rangle}
\newcommand{\pp}{+ \!\!\! +}
\newcommand{\mm}{=}
\newcommand{\qq}{{q \over 2}}

\def\ZZ{Z\!\!\!Z} 		%%%%%%%%%%%%%
\newcommand{\ts}{\textstyle}

\begin{document}
\baselineskip=17pt

\begin{titlepage}

\hfill {USC-94/9}

\hfill {PUPT-1469}

\hfill {hep-th/9406020}

\vskip 1.0cm
\begin{center}

{\Large Spinon Bases, Yangian Symmetry and}\\
\vspace{4mm}
{\Large Fermionic Representations of Virasoro Characters }
\vspace{4mm}
{\Large in Conformal Field Theory}

\vskip 1.8cm

{\large Peter Bouwknegt}

\vskip .3cm

{\sl Department of Physics and Astronomy \\
     University of Southern California \\
     Los Angeles, CA 90089-0484}

\vskip 1.3cm

{\large Andreas W.W. Ludwig, Kareljan Schoutens}

\vskip .3cm

{\sl Joseph Henry Laboratories, Princeton University \\
     Princeton, NJ 08544, U.S.A.}

\vskip .7cm

{\bf Abstract}
\end{center}

\baselineskip=17pt

We study the description of the $SU(2)$, level $k=1$,
Wess-Zumino-Witten conformal field theory in terms
of the  modes of the  spin-1/2 affine primary field $\phi^\alpha$.
These are shown to satisfy generalized `canonical commutation
relations', which we use to construct a basis of Hilbert space
in terms of representations of the
Yangian $Y(sl_2)$. Using this description,
we explicitly derive so-called  `fermionic
representations' of the Virasoro characters, which were
first conjectured by Kedem et al.~\cite{kedem}. We point out
that similar results are expected for a wide class of
rational conformal field theories.

\vspace{.2cm}

\noindent
\baselineskip=17pt

\vfill

\noindent May 1994
\end{titlepage}

\newpage

\baselineskip=17pt

\paragraph{Introduction.}

Recently a number of independent studies
have pointed at a    novel description
 of the structure of the Hilbert
space of certain rational conformal field theories (RCFT),
which is totally different from the conventional one in
terms of representations of a chiral conformal algebra:
roughly speaking, the Hilbert space
can be built up from fundamental `quasiparticles,'
in a way that is  reminiscent
of the Fock space construction for free fermions.  A rule, generalizing
the Pauli exclusion principle, governs the allowed ways
the `quasiparticles' can occupy  single momentum states.
Clearly, such a novel description of RCFT is expected to
lead to a  number of  new insights, both at the mathematical level
and in applications.
Early results in this spirit can be found in the work
of Faddeev and Takhtajan~\cite{fadtak}, and of Zamolodchikov and
Zamolodchikov~\cite{zam-zam}.

In this Letter, we make a connection between
two developments in this area.
 The first goes back to
Haldane et al.~\cite{hhtbp},
who propose a description
of the $SU(2)$, {\mbox level-1} Wess-Zumino-Witten (WZW)
conformal field theory in terms of `{\it spinons}'
 (spin-1/2 doublets)
 and so-called  {\it Yangian symmetry}.
This description, which has its origin in the
structure of the Haldane-Shastry spin chains with $1/r^2$
exchange, stresses the fact that the fundamental
fields in this theory are `spinon'  fields, which may be viewed
as free fields apart from purely statistical
(in this case: semionic) interactions that may be taken into account
by a rule generalizing the Pauli principle.
The second development, initiated by the Stony Brook
group~\cite{kedem}, is the observation that the fundamental
characters of many RCFT's can be written in what has been called
a {\it `fermionic representation'}.
So far, the `fermionic representations' have not been
related to an underlying algebraic structure in the CFT.

Here, following \cite{hhtbp},
we construct a basis of the Hilbert space
of the $SU(2)_1$ WZW model in terms of modes of the spin-1/2 affine
primary field  and relate it to representations of the
Yangian $Y(sl_2)$~\cite{bernard-comment}.
In the main part of this Letter,
we shall then use this spinon description to directly derive
the `fermionic representations'
of the Virasoro characters in the
model. We thus identify the common structure underlying
both developments mentioned above.
We will be able to give
a straightforward derivation, which uses little more
than the generalized commutation relations
of the spinon modes.

In our closing paragraph, commenting
on our results, we argue
that, if a RCFT can be described by different (bosonic)
chiral algebras, one expects a `fermionic representation'
of the characters and a generalized notion of `Yangian symmetry'
for each of those descriptions. We also comment on the
$SU(2)$ WZW models with level $k>1$.

\paragraph{Spinon fields: OPE's and generalized commutators.}

In this paragraph we begin our discussion of the $SU(2)_1$
 WZW model by specifying the properties of the
modes of the (chiral) spinon field $\phi^{\pm}(z)$, which
is nothing else than the $j=\half$ primary field of the
affine symmetry algebra $A^{(1)}_1$. If one were to base
the description of the theory on affine symmetry alone,
one would have the vacuum and the lowest spinon state
(of conformal dimension $L_0=\quart$) as the fundamental
primary states, and one would use affine currents to generate
all other states in the theory. Here we shall focus instead
on a description where all states in the spectrum are
written as multi-spinon states, \ie, as products
of  modes of the spin-1/2 affine
primary.

In order to clarify conventions, we give a complete
set of OPE's. The current algebra is
\be
J^a(z)J^b(w) = {d^{ab} \over (z-w)^2}
               + {f^{ab}{}_c J^c(w) \over (z-w)}
               + \ldots
\ee
The adjoint index  takes the values
$a=\pp,3,\mm$; the metric
is $d^{\pp\mm}= 1$, $d_{\mm\pp} = 1$, $d^{33} =2$,
$d_{33}= \half$ and
the structure constants follow from $f^{\pp\mm}{}_{3}= 1$
and satisfy $f_a{}^{bc} f_{dbc} = -4 \, d_{ad}$.
The spinon fields $\phi^\alpha(z)$, $\alpha=\pm$, transform as
\be
J^a(z)\phi^\alpha(w) = (t^a)^\alpha{}_\beta
                       {\phi^\beta(w) \over (z-w)}
                       + \ldots \ ,
\ee
with $(t^{\pp})^-{}_+ = (t^{\mm})^+{}_- = 1$,
$(t^3)^\pm{}_\pm =  \pm 1$.
Furthermore, we have the following spinon-spinon OPE's
\bea
\phi^\alpha(z)\phi^\beta(w) &=&
    (-1)^q (z-w)^{-\half} \epsilon^{\alpha\beta}
    \left( 1 + \half (z-w)^2 T(w) + \ldots \right)+
\nonu
&& + (-1)^q (z-w)^{\half} (t_a)^{\alpha\beta}
    \left( J^a(w) + \half (z-w) \del J^a(w) + \ldots \right) \ ,
\eea
where $\epsilon^{+-} = -\epsilon^{-+}= 1$,
and $(t_a)^{\alpha\beta} = d_{ab}(t^b)^\alpha{}_\gamma
\epsilon^{\gamma\beta}$.
In these formulas, $q$ depends on the sector that the
OPE's are acting on~\cite{q-dependence}: $q=0$ on
states that are created
by an even number of spinons (\ie, the states in the
vacuum module of $A_1^{(1)}$)
and $q=1$ if the number of spinons is odd (the $j=\half$
module of $A_1^{(1)}$).

The occurence of a factor $(z-w)^{1/2}$ in the
spinon-spinon OPE's clearly shows that the braiding properties
of the spinons are those of semions or `half-fermions.'
The mode expansions of the spinons  are
\bea
\phi^{\alpha}(z)\chi_q(0) = \sum_m z^{m+\qq}
\phi^{\alpha}_{-m-\qq-\quart}
  \chi_q(0) \ ,
\nonu
\phi^{\alpha}_{-m-\qq+{3 \over 4}} \chi_q(0) =
\oint {dz \over 2 \pi i} z^{-m-\qq}
\phi^{\alpha}(z)\chi_q(0) \ ,
\eea
where $\chi_q(0)$ is an arbitrary state in the sector indicated
by the value of $q$. On states with $q=0$ we can apply modes
$\phi^\pm_{-1/4-n}$ with $n$ integer, and on states with
$q=1$ we can apply $\phi^\pm_{-3/4-n}$.

By following a standard procedure (see \eg\ \cite{faza})
we can derive the following {\it generalized commutation
relations}\ for the modes of the spinon fields
\bea
&& \sum_{l\geq0} C_l^{(-{1 \over 2})}
   \left( \phi^\alpha_{-m-{q+1 \over 2}-l+{3 \over 4}}
          \phi^\beta_{-n-\qq+l+{3 \over 4}}
   - \left(
\begin{array}{c}
 \alpha  \leftrightarrow \beta
\\
m \leftrightarrow n
\end{array} \right) \right)
\nonu
&& \qquad \qquad
   = (-1)^q \, \epsilon^{\alpha\beta} \, \delta_{m+n+q-1} \ ,
\label{gc1}
\\[2mm]
&& \sum_{l\geq0} C_l^{(-{3 \over 2})}
   \left( \phi^{\alpha}_{-m-{q+1 \over 2}-l-{1 \over 4}}
          \phi^{\beta}_{-n-\qq+l+{3 \over 4}}
   + \left(
\begin{array}{c}
 \alpha  \leftrightarrow  \beta
\\
 m \leftrightarrow n
\end{array} \right) \right)
\nonu
&& \qquad \qquad = (-1)^q \left( - \epsilon^{\alpha\beta} (m+\qq)
   \delta_{m+n+q} + (t_a)^{\alpha\beta} J^a_{-m-n-q} \right) \ ,
\label{gc2}
\\[2mm]
&& \sum_{l\geq0} C_l^{(-{5 \over 2})}
   \left( \phi^\alpha_{-m-{q+1 \over 2}-l-{5 \over 4}}
          \phi^\beta_{-n-\qq+l+{3 \over 4}}
   - \left(
 \begin{array}{c}
\alpha  \leftrightarrow \beta
\\
m \leftrightarrow n
 \end{array} \right) \right)
\nonu
&& \qquad\qquad = (-1)^q \left( \half \,
     \epsilon^{\alpha\beta} \,
     (m+\qq)(m+1+\qq) \delta_{m+n+q+1} \right.
\nonu
&& \qquad\qquad\qquad\qquad \left.
 + \half (t_a)^{\alpha\beta} (n-m) J^a_{-m-n-q-1}
 + \half \, \epsilon^{\alpha\beta} L_{-m-n-q-1} \right) \ .
\label{gc3}
\eea
In these relations, the coefficients $C_l^{(\alpha)}$ are defined
by the expansion
\be
(1-x)^\alpha = \sum_{l\geq 0} C_l^{(\alpha)} x^l \ .
\ee
The relations (\ref{gc1}) can be interpreted
as {\it generalized canonical commutation relations} of the
fundamental spinon fields. The other relations can be used to
express the current modes $J^a_n$ and the Virasoro generators
$L_n$ as bilinears in spinon modes. We would like to stress
that, up to the complication of the infinite series in the
mode index $l$, these relations are very reminiscent of the
anticommutation relations for free fermions and of the
formulas that express affine and Virasoro currents as fermion
bilinears.

\paragraph{Yangian symmetry.}

We recall from \cite{hhtbp},
that the $SU(2)_1$ WZW  RCFT
provides a realization of the Yangian $Y(sl_2)$.
The Yangian  generators
are
\be
Q_0^a = J_0^a \, , \quad Q_1^a = \half f^a{}_{bc} \sum_{m>0}
  J_{-m}^b J_m^c \ .
\label{yn}
\ee
It is easy to see that the Virasoro generator $L_0$ commutes
with the generators in (\ref{yn}).
When acting on integrable highest weight representations
of $A^{(1)}_1$ at level-1, these generators satisfy the
defining relation of Drinfel'd's Yangian (\cite{drin}).
The Yangian is endowed with a {\it comultiplication} $\Delta$, which
can be viewed as a generalization of angular momentum coupling.
When the action of the Yangian is defined on two
spaces, it acts on the tensor product as follows:
\be
\Delta(Q^a_0)
= {\bf 1} \otimes Q^a_0 +
 Q^a_0  \otimes {\bf 1}, \qquad
\Delta(Q^a_1)
= {\bf 1} \otimes Q^a_1 +  Q^a_1  \otimes {\bf 1}
  + \half f^a{}_{bc}  Q^b_0  \otimes  Q^c_0 \ .
\label{coproduct}
\ee

The paper \cite{hhtbp} also identified a number of
operators (beyond the Virasoro zero-mode $L_0$)
that commute with the Yangian generators. The first
non-trivial example is the operator $H_2$ given by
\be
H_2 = \sum_{m>0} d_{ab} \, m \, J^a_{-m} J^b_m \ .
\label{H2}
\ee

If the group $SU(2)$
is replaced by $SU(N)$, the corresponding expressions
for $Q_1^a$ and $H_2$ have to be modified by additional
terms that involve the 3-index $d$-symbol of $SU(N)$,
$N\geq 3$~\cite{kjs}.

\paragraph{Constructing the Hilbert space.}

We now come to the description of the full space of
states of the theory in terms of multi-spinon states.
There are actually two ways to set this up, and we shall
discuss both in this paragraph.

\subparagraph{Basis I.}

Rephrasing the prescription  of \cite{hhtbp},
we claim that a complete basis for all the (chiral) states in
the Hilbert space of the $SU(2)_1$ WZW model
is given as follows.
 We first construct the following {\it fully
polarized $N$-spinon states}
\bea
&&    \phi^+_{-{(2N-1) \over 4}-n_N}  \ldots
      \phi^+_{-{5 \over 4}     -n_3}
      \phi^+_{-{3 \over 4}     -n_2}
      \phi^+_{-{1 \over 4}     -n_1} |0\rangle \, ,
\nonu
&& \qquad {\rm with} \;\; n_N \geq n_{N-1} \geq \ldots
    \geq n_2 \geq n_1 \geq 0 \ .
\label{pol}
\eea
It is easily seen that the eigenvalue of the
Virasoro zero mode $L_0$ on these states is
\be
   L_0 = {N^2 \over 4} + \sum_{i=1}^N n_i \ .
\label{Lzero}
\ee

In the second step we construct a collection of irreducible
{\it Yangian multiplets}: we first construct a Yangian highest
weight state (YHWS) [annihilated by $Q_0^{\pp}$ and
$Q_1^{\pp}$]  by taking suitable linear combinations of
the states in (\ref{pol}), with fixed $N$ (number of `spinons')
and fixed $L_0$ eigenvalue (see \eg\ (\ref{2spYhws})).
Then we repeatedly apply the generators $Q_0^a$ and $Q_1^a$.
The union of all these Yangian multiplets precisely forms
a basis for the Hilbert space.

The structure of the irreducible Yangian multiplets,
described in \cite{hhtbp}, following \cite{cp},  reads
in our present language as follows.
(i): Each Yangian multiplet is characterized by a set of
non-decreasing integers
${ \{ n_i \}}_{i=1,...,N}$
as in
 (\ref{pol}).
 (ii): The eigenvalue  of $L_0$ ( commuting
with the Yangian ) on the
states in the Yangian multiplet specified by
${ \{ n_i \}}_{i=1,...,N}$  is given by (\ref{Lzero}).
(iii): When acting on a Yangian highest weight
state (which  will be a linear combination of
`fully polarized' states of the form (\ref{pol})),
the Yangian generators (\ref{yn})
create states of a similar form, which however
have some of the
$+$ indices replaced by $-$, and which have different
coefficients in the linear combination. If we were to
act only with $Q_0^a$ we would find a total of $N+1$ such states;
the maximal possible number when acting with the full Yangian
is $2^N$. This maximal number is only realized if
the mode-indices $n_i$ are all different. If some of the $n_i$'s
are equal, the corresponding product
of doublets is projected on the symmetric combination. For
example, a 2-spinon Yangian multiplet will have $3+1=4$ states
if $n_2\neq n_1$, but only 3 states if $n_2=n_1$. ( For comparison,
note that for free spinful fermions, the Pauli-principle
forces 2 fermions with equal momentum to be
in a  singlet state, the triplet
being dropped. In this sense the Yangian generalizes
the Pauli-principle.)

In \cite{hhtbp}, the above description of the space
of states of the WZW model was obtained as an extrapolation
of results based on the exact solution of so-called
Haldane-Shastry spin chains. However, we can easily see
that this result is a rather direct consequence of the
generalized commutation relations given above, and in
particular of the relations (\ref{gc1}). Using this
relation with $\alpha=+$, $\beta=+$, one can show that
each of the $\phi^+$ modes can be applied only once,
and that mode-indices $n_1,n_2,\ldots$ of the state
(\ref{pol}) can be chosen in a preferred order. The
relation (\ref{gc1}) with $\alpha=+$, $\beta=-$ can be
used to show the existence of null states which reduce
the number of states in the Yangian multiplets. We will
further illustrate this in the next paragraph.

\subparagraph{Basis II.}

A second way to write a  multi-spinon basis for the
$SU(2)_1$ WZW model is as follows. One considers
the states
\bea
&&    \phi^-_{-{2(N^+ + N^-)-1 \over 4} -n^-_{N^-}}
      \ldots
      \phi^-_{-{2(N^+ +1)-1 \over 4}    -n^-_1}
      \phi^+_{-{2N^+ -1 \over 4}        -n^+_{N^+}}
      \ldots
      \phi^+_{-{1 \over 4}-n^+_1} |0\rangle \, ,
\nonumber\\[4mm]
&& \qquad {\rm with} \;\;
    n^+_{N^+} \geq \ldots \geq n^+_2 \geq n^+_1 \geq 0
    \ , \quad
    n^-_{N^-} \geq \ldots \geq n^-_2 \geq n^-_1 \geq 0 \ .
\label{plusmin}
\eea
Once again, it can easily be checked that the generalized
commutation relations (\ref{gc1}) allow one
to write every mixed-index multispinon state as a sum of
states of the form (\ref{plusmin}). The eigenvalue of $L_0$
is now given as
\be
   L_0 = { (N^+ + N^-)^2 \over 4}
         + \sum_{i=1}^{N^+} n^+_i
         + \sum_{i=1}^{N^-} n^-_i \ .
\ee

\paragraph{The action of $Q_1^a$ and $H_2$ on
 $N$-spinon states.}

In this paragraph we study the explicit action of the
Yangian generators $Q_0^a$ and $Q_1^a$ and of the
operator $H_2$ on multi-spinon states. For simplicity
we shall first discuss the action on two-spinon states
and later give some more general formulas.

We introduce the following notations for general
2-spinon states ($t$, $s$ refer to triplet and
singlet states, respectively)
\begin{equation}
  \Phi_{n_2,n_1}^{t,a} =
  (t^a)_{\alpha\beta}  \ \phi^\alpha_{-3/4-n_2}
                         \phi^\beta_{-1/4-n_1} \vac,
\quad
  \Phi_{n_2,n_1}^s =
  \epsilon_{\alpha\beta} \ \phi^\alpha_{-3/4-n_2}
                            \phi^\beta_{-1/4-n_1} \vac \ .
\end{equation}
We can now use
\be
   (J^a\phi^\alpha)(z) = 2
          (t^a)^\alpha{}_\beta \del \phi^\beta \ ,
\qquad
   (\del J^a\phi^\alpha)(z) = {4 \over 3}
          (t^a)^\alpha{}_\beta \del^2 \phi^\beta
\label{nfs}
\ee
\noindent to show that
\bea
&& Q_1^a \Phi_{n_2,n_1}^{t,b} =
  -(n_2+n_1+\half) f^{ab}{}_c \Phi_{n_2,n_1}^{t,c}
  + (n_2-n_1+1) d^{ab} \Phi_{n_2,n_1}^s
    + d^{ab} \sum_{l>0}  \Phi_{n_2+l,n_1-l}^s
\nonu
&& Q_1^a \Phi_{n_2,n_1}^s =
  2 (n_2-n_1) \Phi_{n_2,n_1}^{t,a}
  - 2 \sum_{l>0} \Phi_{n_2+l,n_1-l}^{t,a} \ .
\eea
Notice that the action of $Q_1^a$ is not diagonal
in the indices $(n_2,n_1)$ but rather lower-triangular
in the sense that $(n_2,n_1)$ gets mapped into
$(n_2+l,n_1-l)$ with $l\geq 0$ and $n_1 - l \geq 0$.

{}From the action of $Q_1^{\mm}$ it is easily seen that
the space of all two-spinon states with $n_1+n_2=n$ fixed
can be decomposed into multiplets of the
Yangian, whose highest weight states are of the form
\be
\Phi_{n_2,n_1}^{t,\pp} + \sum_{l>0} a_{n_2,n_1}^{(l)}
    \Phi_{n_2+l,n_1-l}^{t,\pp} \ ,
\label{2spYhws}
\ee
where the $a_{n_2,n_1}^{(l)}$ are real coefficients.

These multiplets contain a
triplet and a singlet of $SU(2)$, \ie, a total of four
states, except if $n_1=n_2$, when
the relation (\ref{gc1}) can be used to show
that the singlet is absent.

We remark that  $Q_1^a$  acts by comultiplication
(\ref{coproduct}) on the 2-spinon YHWS, given
its action on the 1-spinon states (which are YHWS).

The action of $H_2$ on the 2-spinon states can be evaluated
in a similar fashion. The result is
\bea
H_2 \Phi_{n_2,n_1}^{t,a} &=&
 2\left( (n_2+1)(n_2+\half) + (n_1+\half)n_1\right)
    \Phi_{n_2,n_1}^{t,a}
 + \sum_{l>0} \, l \, \Phi_{n_2+l,n_1-l}^{t,a} \ ,
\nonu
H_2 \Phi_{n_2,n_1}^s &=&
 2\left( (n_2+1)(n_2+\half) + (n_1+\half)n_1\right)
    \Phi_{n_2,n_1}^s
 - 3 \, \sum_{l>0} \, l \, \Phi_{n_2+l,n_1-l}^s \ .
\eea
Since this  action is again lower triangular, the
eigenvalues of the operator $H_2$ are immediately
seen to equal
$2\left( (n_2+1)(n_2+\half) + (n_1+\half)n_1\right)$.
It can be checked that these values are in agreement
with the prescription given in \cite{hhtbp}, and
also with the formula that was recently given in
\cite{bps}.

To be completely explicit, we present the example
where $n_1+n_2=4$, which are the 2-spinon states with
$L_0=5$ ( from (\ref{Lzero})).
In the following formula we list the labels $(n_2,n_1)$
of the Yangian representation, the $H_2$-eigenvalues, and
the states
\bea
&& (4,0) \qquad H_2=45 \qquad
   \Phi_{4,0}^{t,a}\ , \quad \Phi_{4,0}^s
\nonu
&& (3,1) \qquad H_2=31 \qquad
   \Phi_{3,1}^{t,a} - \ts{\frac{1}{14}} \Phi_{4,0}^{t,a}\ , \quad
   \Phi_{3,1}^s + \ts{\frac{3}{14}} \Phi_{4,0}^s
\nonu
&& (2,2) \qquad H_2=25 \qquad
  \Phi_{2,2}^{t,a} - \ts{\frac{1}{6}} \Phi_{3,1}^{t,a}
       - \ts{\frac{11}{120}} \Phi_{4,0}^{t,a} \ .
\eea
When acting on the $(2,2)$ Yangian highest weight state,
$Q_1^{\mm}$ produces a multiple of the $SU(2)$ descendant
plus a state proportional to
$\Phi_{2,2}^s+ \half \Phi_{3,1}^s + \frac{3}{8} \Phi_{2,2}^s$,
which vanishes as a consequence of (\ref{gc1}).

To conclude this paragraph, we turn to the action
of $Q_1^a$ and $H_2$ on a general $N$-spinon state.
It is easily seen that the action of $Q_1^a$
is again `lower triangular' with respect to a natural
ordering of the multi-indices $(n_1,n_2,\ldots n_N)$.
This implies that in general the space of $N$-spinon
states can be decomposed into irreducible multiplets
of the Yangian. Closed form expressions for general
Yangian highest weight states can \eg\ be obtained by using
the full machinery of the exact solution of the
Calogero-Sutherland models of $N$-body quantum mechanics
and of the Haldane-Shastry spin chains, and this
leads to expressions involving so-called Jack
polynomials~\cite{bps}.

Let us give an explicit expression for the action
of $H_2$ on a fully polarized $N$-spinon state.
Denoting by $|\chi^{(N-1)}\rangle$ a fully
polarized $(N-1)$-spinon state, we have
from (\ref{H2},\ref{nfs})
\bea
\lefteqn{ [ H_2, \phi^+_{-{2N-1 \over 4}-n_N} ]
    |\chi^{(N-1)}\rangle =}
\nonu && \!\!\!\!\!
2 \, (n_N+ {\ts {N-1 \over 2}}  )
     (n_N+ {\ts {N \over 2}}  ) ) \,
    \phi^+_{-{2N-1 \over 4}-n_N}
    |\chi^{(N-1)}\rangle
 + \sum_{l>0} \, l \, \phi^+_{-{2N-1 \over 4}-n_N-l} J^3_l
    |\chi^{(N-1)}\rangle  \ .
\eea

Since this action is again `lower triangular,'
the eigenvalue of $H_2$ on a general fully
polarized Yangian highest weight state with
mode-indices $\{n_1,n_2,\ldots,n_N\}$ is
\be
  H_2 = \sum_{i=1}^N 2(n_i+\half(i-1))(n_i+\half i) \ .
\ee
This value agrees with the results of \cite{hhtbp,bps},
which were derived by different means.

\paragraph{Deriving the Virasoro characters.}

We now turn to a derivation of the Virasoro characters
in the theory by using the characterization of the Hilbert
space that we called `Basis I.' The Virasoro primary
fields in the theory are characterized by their $SU(2)$
spin $j$ and their conformal dimension $\Delta=j^2$.
The corresponding characters are of course well known
\be
\chi^{Vir}_{j^2}(q)
   = {q^{j^2}( 1 - q^{2j+1}) \over
     \prod_{m=1}^\infty (1-q^m)} \ .
\label{viraj}
\ee
Our goal here is to write this character in a way that is
natural from the point of view of the spinon picture.
We shall illustrate our approach by first computing
the vacuum character, $j=0$, which is the generating
function of all $SU(2)$ singlets in the spectrum.

Before proceeding, we introduce a slightly simplified
version of the {\it motif}\ notation~\cite{hhtbp} for
Yangian multiplets as follows
(simplified in the sense that we will not explicitly
write the singlet motif `$(1)$'). Starting with $n_1$, we
write the symbol $()$ for each $n_i$ that is not equal to
one of its neighbours. If a string of $l$ consecutive $n_i$'s
are all equal we write the symbol $(00\ldots0)$
(\ie, $l-1$ zeros).
The Yangian highest weight states are thus in 1-1
correspondence with {\it motif sequences}\ and the $SU(2)$
content of each multiplet is the free tensor product of the
motifs in the sequence.

For example, the Yangian multiplet specified by
$\{n_1 \!= \!n_2 < n_3 < n_4 \!=\! n_5 \!=\! n_6 < n_7\}$
($N=7$ spinons) has motif $ (0)()(00)()$.
Note that the motif notation
only indicates which of the $n_i$'s coincide---there is an
infinite number of Yangian multiplets which have the same motif,
differing by the actual values of a the mode-indices
$n_1<n_3<n_4<n_7$. However, the motif uniquely specifies
the $SU(2)$ content: $()$ has spin $\half$,
$(0)$ spin 1, $(00)$ spin $\threehalves$, etc., and the total
$SU(2)$ content is the free tensor product, for the above
motif $1 \otimes \half \otimes \threehalves \otimes \half$.

Our strategy for the computation of the vacuum character is
as follows. Clearly, this character will only get contributions
from sectors with an even number $m_1$ of spinons.
To find the contribution from the $m_1$-spinon sector,
we proceed in two  steps.
(1): We draw all path (Bratteli) diagrams, which encode possible
ways to extract a singlet from the $m_1$-fold  tensor
product of the $SU(2)$ doublet.  (2): To each diagram we
then associate all allowed motif-sequences from which
a singlet can be extracted according to the diagram
and we sum the corresponding $q$-series.

Starting with step (2), let us assume that we
have a specific Bratteli diagram, which we can also view as a
sequence of uparrows (uaw's) and downarrows (daw's).
To this we associate a {\it leading motif sequence}\
according to the following rule. First write $(00...0)$
[\ie,  $(l-1)$ zeros] for the first sequence of $l$ uaw's. Then
follow the diagram down (daw's) and up again (uaw's)
until the next top. If this dip (which can be asymmetric)
has a total of $l$ arrows, write $(00...0)$ with $l-1$
zeros. Finally, write $(00...0)$ [\ie, $l-1$ zeros] for the final
$l$ daw's. For example, the diagram with first $l$ uaw's
and then $l$ daw's gives $(00...0)(00...0)$. The diagram
with $l$ times the pattern uaw, daw, gives $()(0)(0)...(0)()$
with $l-1$ times $(0)$ in the middle, etc.

{}From the leading motif sequence we construct {\it fragmented
sequences}\ by making the replacement `$0 \rightarrow )($'
in all possible places.

All this is easily explained in an example. Pick the
Bratteli diagram [step (2)]
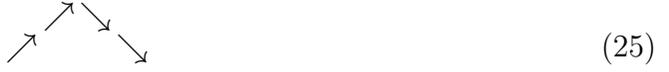
\be
   \nearrow \,
   \stackrel{\displaystyle{\nearrow}}{\rule[0mm]{0mm}{3mm}} \,
   \stackrel{\displaystyle{\searrow}}{\rule[0mm]{0mm}{3mm}} \,
   \searrow
\label{4diagram}
\ee
describing $m_1=4$ spinons. (The diagram starts and ends
at $j=0$, appropriate for the vacuum character.)
The leading motif sequence for this diagram is
$(0)(0)$ and the  possible fragmentations are $()()(0)$,
$(0)()()$ and $()()()()$.

Recall that the motif sequences only
indicate whether some of the $n_i$ in (\ref{pol})
coincide or not. This means that to each motif
sequence there corresponds an infinite number of
choices of the labels $\{n_i\}$, which contributes
a certain $q$-series to the character.

In the case of the example one obtains, using
$ L_0 = 4 + \sum_{i=1}^4 n_i$ (from (\ref{Lzero})),
\bea
{\it motif:} && {\it mode} \ {\it integers:}
\qquad \qquad \quad {\it character:}
\nonu
(0)(0) && 0\leq n_1=n_2<n_3=n_4 \qquad
{\ts {q^6 \over (1-q^2)(1-q^4)}}
\nonu
()()(0) && 0 \leq n_1<n_2<n_3=n_4 \qquad
{\ts {q^9\over (1-q^2)(1-q^3)(1-q^4)}}
\nonu
(0)()() && 0 \leq n_1=n_2<n_3<n_4 \qquad
{\ts {q^7 \over (1-q^1)(1-q^2)(1-q^4)}}
\nonu
()()()() && 0 \leq n_1<n_2<n_3<n_4 \qquad
{\ts {q^{10} \over (1-q)(1-q^2)(1-q^3)(1-q^4)}}
\label{characterexample}
\eea
The sum of all four contributions to the character
equals $q^6/(q)_4$, where we use the notation
(for $a \in {\ZZ}_{>0}$)
\be
(q)_a = \prod_{n=1}^a (1-q^n)
\ee
with $(q)_0=1$ and $(q)_{-a}=0$.

It is easy to show that in general all contributions
that correspond to any given Bratteli  diagram always add up to
\be
{q^{L_0({\rm diagram})} \over (q)_{m_1}} \ .
\ee
In this formula $L_0({\rm diagram})$ is defined
to be the lowest $L_0$ value for a state corresponding
to the leading motif sequence of that diagram
[$(0)(0)$ for the example (\ref{4diagram}),
giving $L_0=6$].

Our remaining task is to sum $q^{L_0({\rm diagram})}$
over all possible Bratteli diagrams [step (1)].
Let us start by evaluating
the sum at $q=1$, \ie, by simply counting the number
of such diagrams. It is well known that this number can
be written  in terms of binomial coefficients
\be
\#({\rm singlets}\;{\rm in}\; 2^{m_1}) =
\sum_{m_2,m_3,\ldots\in 2{\ZZ}}\ \prod_{a\geq 2}
\left( \begin{array}{c} \half(m_{a-1}+m_{a+1}) \\ m_a
       \end{array} \right) \ .
\label{sings}
\ee
To a given set
$\{m_1,m_2,\ldots\}$ of even integers corresponds
a set of Bratteli diagrams as follows
(the $m_a$'s must satisfy
$ \half(m_{a-1}+m_{a+1}) \geq m_a$ and only a finite
number of them are non-zero; the non-zero ones are
strictly descending).
Suppose that $m_l$ is the highest $m_i$ not equal to zero.
We start by drawing a pattern
that has first $l$ uaw's, then $l$
daw's, etc, repeating this pattern $\half m_l$ times
($m_l$ is even). Next we insert $(\half m_{l-1}
-m_l)$ times a similar pattern, of lenght $2(l-1)$ instead of
$2l$. In principle, the pattern can be inserted as a `top'
(first uaw's, then daw's) or as a `dip' (first daw's, then
uaw's). The rule is that if the arrow on the left of the
insertion points up  you put a `dip,' else a `top.'
It may be checked  that, since all
allowed diagrams are to be counted precisely once,
there are $m_l+1$ positions where the insertions
(which may be multiple) can be done. This means that we are
separating $m_l$ objects by $(\half m_{l-1} -m_l)$ separators,
and this can be done in
$\left( \begin{array}{c} \half m_{l-1} \\
  m_l \end{array} \right)$
ways. Next, we insert $(\half(m_{l-2} + m_{l})-m_{l-1})$ times
a top or a dip of length $2(l-2)$. There are $m_{l-1}+1$
spots where the insertions can be done and the
corresponding factor is
$ \left( \begin{array}{c} \half (m_{l-2}+m_{l}) \\
  m_{l-1} \end{array} \right)$.
Continuing, we build up the full product in (\ref{sings}).

We should now compute
$L_0$ for each of the diagrams
and sum $q^{L_0}$ over all diagrams associated
with a set $\{m_1,m_2,\ldots\}$. We claim that the result is
\be
q^{\half(m_1^2+m_2^2+\ldots-m_1m_2-m_2m_3-\ldots)}
\prod_{a\geq 2}
\left[ \begin{array}{c} \half(m_{a-1}+m_{a+1}) \\ m_a
       \end{array} \right]_q \ ,
\label{mfixed}
\ee
where
\be
\left[ \begin{array}{c} a \\ b
       \end{array} \right]_q
=
{(q)_a \over (q)_{a-b}(q)_b}, \quad {\rm for}
\ \ a \geq b \ ,
\ee
(and zero otherwise).
This formula can be understood as a $q$-deformation of the
combinatorical expression  in (\ref{sings}), and  it
can be derived by
going through the same steps, this time keeping track of the $q$
dependence of all factors.

As an example, let us choose $m_1=6$,
$m_2=2$ and the rest 0.
Going through the construction of
diagrams as described above ($l=2$ here),
 we first draw the diagram
\be
   \nearrow \,
   \stackrel{\downarrow}{\rule[0mm]{0mm}{3mm}} \,
   \stackrel{\displaystyle{\nearrow}}{\rule[0mm]{0mm}{3mm}} \,
   \stackrel{\displaystyle{\searrow}}{\rule[0mm]{0mm}{3mm}} \,
   \stackrel{\downarrow}{\rule[0mm]{0mm}{3mm}} \,
   \searrow \, \scriptstyle{\downarrow}
\ee
The vertical arrows indicate the positions where, in the
second step, we can insert the pattern $\searrow \nearrow$ or
$\nearrow \searrow$. We thus find a total of
$ \left( \begin{array}{c} 3 \\ 2 \end{array} \right) = 3$
diagrams. With their corresponding leading motif sequence
and $ q^{L_0}$ value they are
\bea
&& \nearrow \,
   \stackrel{\displaystyle{\nearrow}}{\rule[0mm]{0mm}{3mm}} \,
   \stackrel{\displaystyle{\searrow}}{\rule[0mm]{0mm}{3mm}} \,
   \searrow \, \nearrow \, \searrow
   \qquad \quad (0)(00)()  \quad  q^{14}
\nonu
&& \nearrow \,
   \stackrel{\displaystyle{\nearrow}}{\rule[0mm]{0mm}{3mm}} \,
   \stackrel{\displaystyle{\searrow}}{\rule[0mm]{0mm}{3mm}} \,
   \stackrel{\displaystyle{\nearrow}}{\rule[0mm]{0mm}{3mm}} \,
   \stackrel{\displaystyle{\searrow}}{\rule[0mm]{0mm}{3mm}} \,
   \searrow
\qquad \quad (0)(0)(0)  \quad  q^{15}
\nonu
&& \nearrow \, \searrow \,
   \nearrow \,
   \stackrel{\displaystyle{\nearrow}}{\rule[0mm]{0mm}{3mm}} \,
   \stackrel{\displaystyle{\searrow}}{\rule[0mm]{0mm}{3mm}} \,
   \searrow
\qquad \quad ()(00)(0)  \quad  q^{16} \ .
\eea
These add up to $q^{14}(1+q+q^2)$ which is indeed equal to
$q^{14} \left[ \begin{array}{c} 3 \\ 2 \end{array} \right]_q$.

Putting together all ingredients, we find the following
result for the Virasoro vacuum character
\be
\chi^{Vir}_0 = \sum_{m_1,m_2,\ldots \in 2{\ZZ}}
q^{\half(m_1^2+m_2^2+\ldots-m_1m_2-m_2m_3-\ldots)}
{1 \over (q)_{m_1}} \prod_{a\geq 2}
\left[ \begin{array}{c} \half(m_{a-1}+m_{a+1}) \\ m_a
       \end{array} \right]_q \ .
\label{vir0}
\ee
This expression is the $c=1$ limit of a `fermionic
representation' of the Virasoro characters for unitary minimal
models, as first given by Kedem et al.~\cite{kedem}
and proven in \cite{ber}.

Note that the sum over $m_2,m_3,...$ at fixed $m_1$
gives the contribution of the $m_1$-spinon states to
the character.

The Virasoro characters corresponding to non-zero $SU(2)$
spin $j$ can be obtained in an identical way, this time using
Bratteli diagrams that start at $j=0$ and arrive at $j$.
The character formula obtained reads as follows

\hfill \parbox{14.5cm}{
\bea
\chi^{Vir}_{j^2} &=& q^{-j/2} \, \sum^*_{m_1,m_2,\ldots}
q^{\half(m_1^2+m_2^2+\ldots-m_1m_2-m_2m_3-\ldots)}
\nonu
&& \times {1 \over (q)_{m_1}} \prod_{a\geq 2}
\left[ \begin{array}{c} \half(m_{a-1}+m_{a+1}+\delta_{a,2j+1})
       \\ m_a \end{array} \right]_q \ ,
\label{virj}
\eea}

\noindent where the $*$ on the summation symbol
indicates that
$m_1,m_3, \ldots, m_{2j-1}, m_{2j+1}, m_{2j}, \ldots$
are even and $m_2,m_4, \ldots, m_{2j}$ are odd.

\paragraph{$A^{(1)}_1$ characters and generalizations.}

{}From the basis of states which we called Basis II,
it immediately follows that the (level-1) affine characters
that occur in this theory have the following
fermionic representation
\be
\chi^{A^{(1)}_1}_{j=0} =
 \sum_{n^+ + n^- \; {\rm even}}
 {q^{(n^+ + n^-)^2/4} \over (q)_{n^+} (q)_{n^-}} \ ,
\qquad
\chi^{A^{(1)}_1}_{j={1 \over 2}} =
 \sum_{n^+ + n^- \; {\rm odd}}
 {q^{(n^+ + n^-)^2/4} \over (q)_{n^+} (q)_{n^-}} \ .
\ee
These formulas were first written in \cite{ezer} and
they were related to the spinon picture in \cite{bps}.

We thus see that for both choices of chiral algebra in
this CFT (Virasoro and $A^{(1)}_1$) the characters have
a fermionic representation, and that in both cases we can
derive this from the generalized commutation relations
of the fundamental spinon fields.

Looking at other RCFT's, we make a number of observations.
First of all, it seems clear that in general, the
possibility to write `fermionic representations'
of characters in a RCFT directly points at a
specific chiral algebra for which those characters
are appropriate. (McCoy has put forward interesting
speculations on the connection of this with the possibility
to find integrable massive perturbations of the
CFT~\cite{barry}.) Secondly, there is a rather clear
picture of how such a description changes if we pass
from a WZW model to corresponding (diagonal) coset
conformal field theories: in the character formulas
such as (\ref{vir0}) for the $n$-th minimal coset model
only a finite number of labels $m_1,m_2,\ldots,m_n$
are allowed, which implies that the coset CFT's can
be viewed as {\it interacting}\ spinon theories in
which some of the multi-spinon states are lost.

In the language of the Bethe Ansatz equations, the
labels $\{m_2,m_3,\ldots\}$ in (\ref{vir0}) refer to so-called
`ghost' excitations, which do not have a macroscopic range
for their momenta. This is in contrast to $m_1$, which
corresponds to a true quasi-particle. We can now describe
the role of the Yangian symmetry in the $SU(2)_1$
WZW model as follows: starting from the states generated by
the single true quasi-particle (which we take to be
$\phi^+$), the Yangian adds `descendant states,' which in
the Bethe Ansatz language would be associated with the
`ghost' degrees of freedom. In this formulation, the Yangian
symmetry is characterized in an operational way, and one
can try to find its analogue in other RCFT's.

Applying this approach to the level-$2$ $SU(2)$ WZW model,
we observe the following. The character formula for the
$SU(2)$ singlets in the (unprojected)  Neveu-Schwarz
vacuum sector
of the super Virasoro algebra takes
a form~\cite{kedem} which is identical to (\ref{vir0}), with,
however, for a given $m_1$ an additional factor that has
the form
\be
\sum_{m_0 \in {\ZZ}}  q^{{1 \over 2}m_0^2
 - {1 \over 2} m_0 m_1}
 \left[ \begin{array}{c} {1 \over 2} m_1 \\ m_0 \end{array}
 \right]_q   \ .
\ee
(we shifted indices as compared to \cite{kedem}).
This level-2 character clearly suggests
an algebraic structure where, in addition to generators
that are analogous to the Yangian generators for level-1,
there are generators
that create `replicas' of the level-1 Yangian
multiplets. For example, if $m_1=2$ the additional factor
equals $(1+q^{-{1 \over 2}})$, and one
expects a two-fold degeneracy of all two-spinon states.
The details of this algebraic structure, which can be
studied for general level $k>1$, will be published
elsewhere~\cite{bls}.

It is a pleasure to thank P. Argyres, A. Berkovich,
D. Haldane and B. McCoy for illuminating discussions.
The research of KS was supported by DOE grant
DE-AC02-76ER-03072.


\frenchspacing
\baselineskip=16pt
\begin{thebibliography}{11}
\bibitem{fadtak} L.D. Faddeev and L.A. Takhtajan, Phys.
Lett. {\bf 85A} (1981) 375.
\bibitem{zam-zam} A.B. Zamolodchikov and Al.B. Zamolodchikov,
Nucl. Phys. {\bf B379} (1992) 602.
\bibitem{hhtbp}
  F.D.M. Haldane, Z.N.C. Ha, J.C. Talstra, D. Bernard
  and V. Pasquier, Phys. Rev. Lett. {\bf 69} (1992) 2021.
\bibitem{kedem}
  R. Kedem, T. Klassen, B. McCoy and E. Melzer,
  Phys. Lett. {\bf B307} (1993) 68 (and references therein).
\bibitem{bernard-comment}
  Related results were obtained independently
  in Ref.~\cite{bps}.
\bibitem{q-dependence} The appearance of explicit factors
$(-1)^q$ is due to our convention to write upper indices
on all spinon fields. The natural convention would be
to write lower indices whenever a spinon field acts on
a $q=1$ state; raising these indices by using $\phi^\alpha
= \epsilon^{\alpha\beta} \phi_\beta$ leads to an explicit
relative minus sign between the OPE's in the two sectors
$q=0,1$.
\bibitem{faza}
  A.B. Zamolodchikov and V.A. Fateev,
  Sov. Phys. J.E.T.P. {\bf 62} (1985) 215;
  Sov. Phys. J.E.T.P. {\bf 63} (1986) 913.
\bibitem{drin}
  V.G. Drinfel'd, Sov. Math. Dokl. {\bf 32} (1985)
  254; in Proceedings of the International Congress of
  Mathematicians, Berkeley, California (1986).
\bibitem{kjs}
  K. Schoutens, {\it Yangian Symmetry in Conformal Field
  Theory}, PUPT-1442, hep-th/9401154.
\bibitem{cp}
  V. Chari and A. Pressley, L'Enseignement Math. {\bf 36}
  (1990) 267; J. reine angew. Math. {\bf 417} (1991) 87.
\bibitem{bps}
  D. Bernard, V. Pasquier and D. Serban, {\it Spinons in
  Conformal Field Theory}, SPhT/94/039, hep-th/9404050.
\bibitem{ber}
  A. Berkovich, {\it Fermionic counting of $RSOS$-states and
  Virasoro character formulas for the unitary minimal
  series $M(\nu,\nu+1)$. Exact results.},
  preprint BONN-HE-94-04, hep-th/9403073.
\bibitem{ezer}
  E. Melzer, {\it The many faces of a character},
  preprint TAUP 2125-93, hep-th/9312043.
\bibitem{barry}
  B. McCoy, in talk at the conference on `Statistical
  Mechanics and Quantum Field Theory', USC, May 1994.
\bibitem{bls}
  P. Bouwknegt, A. Ludwig and K. Schoutens, in preparation.

\end{thebibliography}
\end{document}